\documentclass[%
 reprint,
 amsmath,amssymb,
 aps,
]{revtex4-2}

\usepackage{textcomp}
\usepackage{graphicx}
\usepackage{dcolumn}
\usepackage{bm}
\usepackage{hyperref}
\usepackage{xcolor}
\usepackage{multirow}
\usepackage{makecell}

\usepackage{tikz, pgfplots}
\pgfplotsset{compat=1.18}
\usepgfplotslibrary{colormaps}
\usepgfplotslibrary{surfshading}
\usepgfplotslibrary{groupplots}
\usepackage{import}
\usepackage{physics}

\def \rcs{Rb$_{2}$Co$_{2}$(SeO$_{3}$)$_{3}$}
\begin{document}

\title{Magnetization Plateaus in the Spin-Orbit Coupled
\\
Bilayer Triangular Lattice Antiferromagnet Rb$_{2}$Co$_{2}$(SeO$_{3}$)$_{3}$}

\author{Shengzhi Zhang}
\altaffiliation{These authors contributed equally to this work.}
\affiliation{National High Magnetic Field Laboratory, Los Alamos National Laboratory, Los Alamos, New Mexico 87545, USA}

\author{Gabriel Silva Freitas}
\affiliation{National High Magnetic Field Laboratory, Los Alamos National Laboratory, Los Alamos, New Mexico 87545, USA}

\author{Wonjune Choi}
\altaffiliation{These authors contributed equally to this work.}
\affiliation{Theoretical Division, T-4 and CNLS,
Los Alamos National Laboratory, Los Alamos, New Mexico 87545, USA}

\author{Shi-Zeng Lin}%
\affiliation{Theoretical Division, T-4 and CNLS,
Los Alamos National Laboratory, Los Alamos, New Mexico 87545, USA}%
\affiliation{Center for Integrated Nanotechnologies (CINT), Los Alamos National Laboratory, Los Alamos, New Mexico 87545, USA}

\author{Tong Chen}
\affiliation{Department of Physics and Astronomy, Johns
Hopkins University, Baltimore, Maryland 21218, United
States}

\author{Xianghan Xu}
\affiliation{Department of Chemistry, Princeton University, Princeton, New Jersey 08544, United States}

\author{Collin Broholm}
\affiliation{Department of Physics and Astronomy, Johns
Hopkins University, Baltimore, Maryland 21218, United
States}
\author{Robert J. Cava}
\affiliation{Department of Chemistry, Princeton University, Princeton, New Jersey 08544, United States}

\author{Eun Sang Choi}
\affiliation{National High Magnetic Field Laboratory, Tallahassee, Florida 32310, USA}

\author{Vivien S. Zapf}%
\affiliation{National High Magnetic Field Laboratory, Los Alamos National Laboratory, Los Alamos, New Mexico 87545, USA}%
\author{Minseong Lee}
\email{ml10k@lanl.gov}
\affiliation{National High Magnetic Field Laboratory, Los Alamos National Laboratory, Los Alamos, New Mexico 87545, USA}

\date{\today}

\begin{abstract}
Geometric frustration among competing spin exchanges can give rise to novel quantum phases by enhancing fluctuations that drive magnetic systems beyond the classical regime.
We investigate the frustrated array of strongly correlated spin dimers in the bilayer triangular lattice antiferromagnet \rcs{} under applied magnetic fields.
A cascade of magnetization plateaus appears at \(M/M_s = 1/3, 1/2, 2/3,\) and \(5/6\), together with a weak anomalous feature near \(M/M_s = 1/6\), in fields up to 60 T.
Concurrent changes in magneto-dielectric response follows the plateau boundaries.
The finite slope of each plateau and the absence of a zero-field gap in our ultralow-temperature ac susceptibility down to 20 mK indicate broken \(U(1)\) spin-rotation symmetry. 
A minimal bilayer-dimer model treated with bond operator representation reproduces the low-field sequence only when \(U(1)\) symmetry is explicitly lifted by spin-orbit-driven, bond-dependent anisotropy.
Near saturation, a projected triangular pseudospin model accounts for the high-field plateaus with modest further-neighbor interactions.
These results demonstrate that anisotropic exchange arising from spin-orbit-coupled moments is essential for stabilizing the full plateau hierarchy in \rcs{}, a mechanism overlooked in previous interpretations of Co-based triangular bilayers.
\end{abstract}

\maketitle


\section{\label{sec:intro}Introduction}
Magnetization plateaus are among the most striking manifestations of quantum many-body effects in magnetic insulators \cite{oshikawa1997magnetization,momoi2000magnetization,kumar2014chern}. Such plateaus often cannot be explained by simple classical spin alignment; instead, they can signal strong correlations, discrete symmetry breaking, or emergent commensurability arising from the interplay of exchange anisotropy, lattice geometry, and quantum or thermal fluctuations \cite{chubukov1991quantum,lee2014magnetic,seabra2011phase}. Beyond their phenomenological interest, magnetization plateaus provide valuable insight into the microscopic spin Hamiltonian. Their occurrence constrains the types of magnetic interactions capable of stabilizing quantized states, while their widths reflect the characteristic energy scales and relative strengths of exchange couplings \cite{karl2020insights}. Consequently, systematic studies of magnetization plateaus offer a powerful means to elucidate the microscopic interactions and emergent symmetries that govern complex magnetic materials.

Following this general understanding, magnetization plateaus are often observed in frustrated magnets, where competing interactions give rise to a rich variety of field-induced quantum phases \cite{zhitomirsky2015real,schluter2022melting}. In systems such as the triangular lattice antiferromagnet \cite{scheie2024spectrum,lee2024magnetic} and the Shastry–Sutherland lattice \cite{jaime2012magnetostriction, kodama2002magnetic}, the origins of these plateaus have been extensively studied and are now relatively well understood. In the triangular lattice, strong frustration gives rise to pronounced quantum fluctuations that stabilize magnetization plateaus of finite width \cite{kamiya2018nature,starykh2015unusual,chubukov1991quantum}, whereas in the Shastry–Sutherland system, both quantum fluctuations and magnetoelastic coupling are known to play essential roles in the formation of multiple plateau phases\cite{jaime2012magnetostriction,kodama2002magnetic}.

Recently, the bilayer triangular-lattice antiferromagnet K$_{2}$Co$_{2}$(SeO$_{3}$)$_{3}$ has drawn significant attention because of a series of magnetization plateaus observed in high-field magnetization measurements \cite{fu2025berezinskii,zhong2020frustrated,li2025differences}. In pulsed-field experiments, researchers reported a sequence of five plateaus at approximately 1/6, 2/6, 3/6, 4/6, and 5/6 of the saturation magnetization, with the 1/6 plateau appearing only weakly, suggesting a fragile or partially stabilized spin configuration. To account for this unusual series, several theoretical models have been proposed \cite{effective_TFIM, xu2024frustrated}. One approach successfully describes the low-field plateaus in terms of spin dimerization and the triangular-lattice symmetry, which can naturally stabilize fractional magnetization states through frustration-induced quantum effects \cite{effective_TFIM}. In contrast, another interpretation based on the local crystal structure suggested that the nearest-neighbor Co–Co interactions may include competing ferromagnetic (FM) and antiferromagnetic (AFM) contributions that nearly cancel each other \cite{xu2024frustrated}. 

In this context, we investigated the magnetic properties of \rcs{} in detail by performing high-field magnetization, ultralow-temperature ac susceptibility, and high-field dielectric constant measurements. We developed a microscopic spin model to provide insight into the multiple magnetization plateaus observed in \rcs{}, which closely resemble those reported in K$_{2}$Co$_{2}$(SeO$_{3}$)$_{3}$ \cite{xu2024frustrated}. Although \rcs{} and K$_{2}$Co$_{2}$(SeO$_{3}$)$_{3}$ are nearly identical in most structural and magnetic aspects, our model assumes that each bilayer is well separated, making the interbilayer coupling negligible. Therefore, \rcs{}—with its larger interbilayer spacing (6.8660 \AA) compared to K$_{2}$Co$_{2}$(SeO$_{3}$)$_{3}$ (6.6176 \AA) while the distances between nearest neighbor Co$^{2+}$ are 2.9434 \AA{} for \rcs{} and 2.9452 \AA{} for K$_{2}$Co$_{2}$(SeO$_{3}$)$_{3}$ —is more suitable for this description.

Taking the bilayer as the magnetic unit, we analyze a $U(1)$-symmetric XXZ model with strong interlayer exchange that forms spin-singlet dimers.
In this model, the ground state has finite spin gap and plateaus are all strictly flat, which disagrees with the measured data.
Adding $U(1)$-symmetry breaking spin exchanges that mix the $S^z=0$ and $S^z=\pm1$ spin-triplet excitations remove the zero-field gap and produces finite slopes on the plateaus.
In the high-field limit we derive a triangular-lattice pseudospin model, where modest further-neighbor couplings can stabilize all experimentally observed high-field plateaus.
Taken together, experiment and theory show that models built only from $U(1)$-symmetric exchanges are insufficient and highlight essential role of bond-dependent anisotropic interactions. 
Such terms arise naturally in spin-orbit-coupled Mott insulators and are prominent in Co-based magnets \cite{jackeli2009mott,liu2018pseudospin,liu2020kitaev,liu2021towards,sano2018kitaev,motome2020materials}, which may play a decisive role in breaking the spin-rotation symmetry and stabilizing the observed plateau phases.

\section{\label{sec:method}Method}

Magnetization as a function of magnetic field was measured using millisecond pulsed magnets up to 60 T at the Pulsed Field Facility of the National High Magnetic Field Laboratory, Los Alamos National Laboratory. A single crystal of \rcs{} was mounted with {\bf {\it H}} ‖ {\it c} inside a nonmagnetic ampoule using Apiezon N grease to ensure thermal contact and mechanical stability. The sample was positioned within a radially counterwound copper pickup coil, which effectively cancels the background voltage induced by the spatially uniform pulsed magnetic field. Because the magnetic moment of the sample is spatially nonuniform on the scale of the coil, the voltage induced by the changing magnetic moment during the field pulse was recorded and subsequently numerically integrated in time to obtain the magnetization. Residual drift arising from thermal contraction of the compensation coil was corrected by subtracting an appropriately scaled signal from an auxiliary loop of wire. An additional background subtraction was performed by measuring with the sample inserted and removed from the coil using an in situ extraction rod. The sample was thermally anchored in either $^{3}$He or $^{4}$He exchange gas, and the temperature was monitored immediately before each pulse using a Cernox thermometer (Lakeshore). The absolute magnetization scale was calibrated against DC magnetization data obtained using a vibrating-sample magnetometer in a 14 T Physical Property Measurement System (Quantum Design).

To measure the magnetocaloric effect (MCE)—the temperature change as a function of magnetic field—an AuGe thin-film thermometer was deposited directly on the sample surface by RF magnetron sputtering at a pressure of 40 mTorr of ultrahigh-purity Ar gas for 60 minutes with 100 W RF power. A reference AuGe film was simultaneously prepared on a glass substrate. The resistances of both sample and reference thermometers were calibrated against a Cernox sensor at zero field. During the pulsed-field experiments, the resistances of the AuGe films were recorded as functions of magnetic field, and the background magnetoresistance was subtracted using the reference thermometer.

To measure the dielectric constant, we employed a recently developed transport method using a virtual-ground configuration \cite{peria2025development}. Silver epoxy was applied to the two large, parallel surfaces of the sample (normal to the {\it c} axis) to form electrical contacts, and gold wires were attached for electrical connection. A TREK voltage amplifier was used to apply an ac voltage of 30 V at 100 kHz to one side of the sample, while the AC current generated on the opposite side was detected using a current-to-voltage amplifier. Further experimental details of this technique can be found in Ref. \cite{peria2025development}.

The AC magnetic susceptibility was measured at SCM1 of the National High Magnetic Field Laboratory in Tallahassee, Florida, using a home-made AC susceptometer. The instrument consists of a solenoidal coil that generates the AC excitation magnetic field and a pair of oppositely wound sensing coils positioned inside the solenoid. The sensing coils are designed to have equal but opposite mutual inductance so that, when connected in series, their induced voltages cancel, resulting in zero net signal in the absence of a sample. We placed a single crystal at the center of one of the sensing coils. The presence of the sample alters the magnetic flux, inducing a net voltage across the sensing coils that is proportional to the ac magnetic susceptibility. For more details, see ref. \cite{lee2016magnetic}. Both the ac excitation field and the external dc magnetic field were applied along the out-of-plane direction. 
\section{\label{sec:ex_Results} Experimental Results}
\begin{figure}
    \centering
    \includegraphics[width=1\columnwidth]{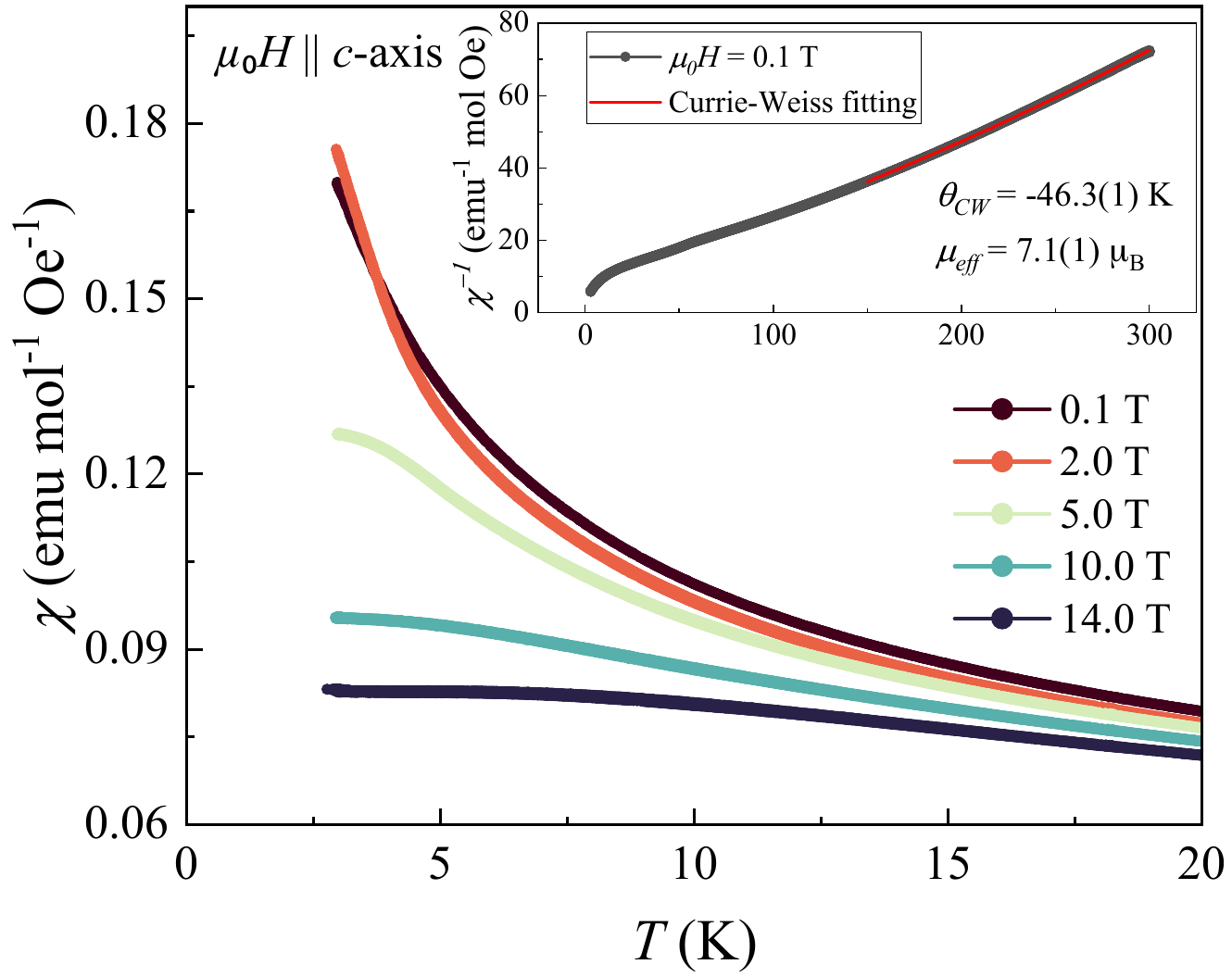}
    \caption{Temperature dependent dc susceptibility $\chi$ at various magnetic fields along $c$-axis.}
    \label{MvsT}
\end{figure}

Fig.~\ref{MvsT} shows the dc magnetic susceptibility measured at various magnetic fields along the c-axis. At 0.1 T, the dc susceptibility exhibits Curie–Weiss behavior between 200 and 300 K, where we fitted the data using the Curie–Weiss formula. We obtained a Curie–Weiss temperature of –46.3(1) K and an effective moment of 7.1(1) $\mu_{B}$, indicating overall antiferromagnetic interactions and a significant contribution from the orbital magnetic moment to the dc susceptibility. At 0.1 T and 2 T, no magnetic ordering was detected, as the susceptibility increases monotonically with decreasing temperature down to 2.5 K. With increasing magnetic field, the curvature of the susceptibility changes, which is a signature of entering the plateau phases, as we discuss later.

the crystals used in this study were grown following the same recipe described in \cite{xu2024frustrated}.

\begin{figure}
    \centering
    \includegraphics[width=1\columnwidth]{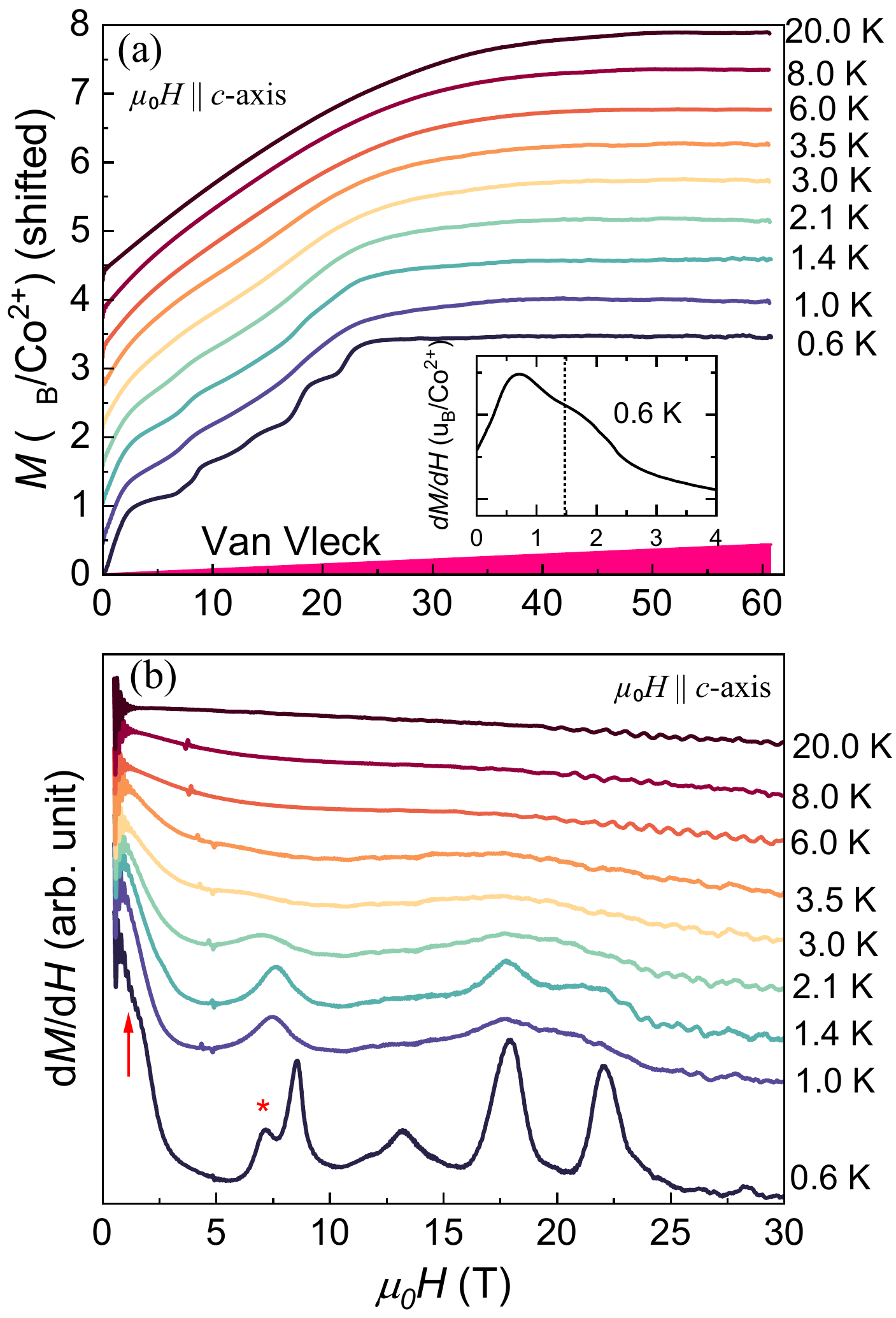}
    \caption{(a) Magnetization measured in pulsed magnetic field at various temperatures. All the curves begin at zero, but vertical offsets were applied for clarity. The Van Vleck shown in the figure was subtracted from the data. (b) The derivatives of magnetization with respect to the magnetic field (d$M$/d$H$). The red arrow indicates the small slope changes around 1/6 plateau consistent with the inset of (a). }
    \label{magnetization}
\end{figure}

Next, we measured the isothermal magnetization as a function of magnetic field along the {\it c}-axis at various temperatures from 0.6 K to 20 K, as shown in Fig.~\ref{magnetization}. A temperature-independent increase in magnetization above 30 T was observed, highlighted as the pink triangular region at the bottom of Fig.\ref{magnetization}. This behavior is attributed to the Van Vleck orbital magnetization, which commonly appears in cobalt-based compounds due to low-lying orbital excitations \cite{susuki2013magnetization}. This contribution was subtracted from all magnetization curves.

At high temperatures above 6 K, the magnetization increases smoothly with field. As the temperature decreases below 6 K, the curvature of the low-field magnetization becomes progressively steeper. Below 1.4 K, distinct magnetization plateaus begin to emerge, becoming clearly resolved into four consecutive plateaus at 0.6 K before reaching full saturation below 24 T, similar to its K$_{2}$Co$_{2}$(SeO$_{3}$)$_{3}$ \cite{fu2025berezinskii}. The fractional values of these plateaus can be assigned as 2/6, 3/6, 4/6, and 5/6 of the saturation magnetization. An additional subtle feature is observed as a small change in slope of the differential susceptibility, marked in the inset of Fig.~\ref{magnetization}, occurring near 1/6 of the saturation magnetization.

The magnetization data suggest a negligible energy gap at zero magnetic field despite the dominant nearest-neighbor exchange interactions, as evidenced by the rapid increase of magnetization at low fields. In addition, the finite slopes of the magnetization plateaus clearly suggest that the spin rotational symmetry along the {\it c}-axis is broken. It is possible that the energy gap is so small that even thermal fluctuations at 0.6 K obscure its signature. To further investigate this, we measured the magnetic ac susceptibility down to 20 mK, enabling us to examine how the magnetization plateaus evolve at ultralow temperatures and if any magnetic long range order develops.

\begin{figure}
    \centering
    \includegraphics[width=1\columnwidth]{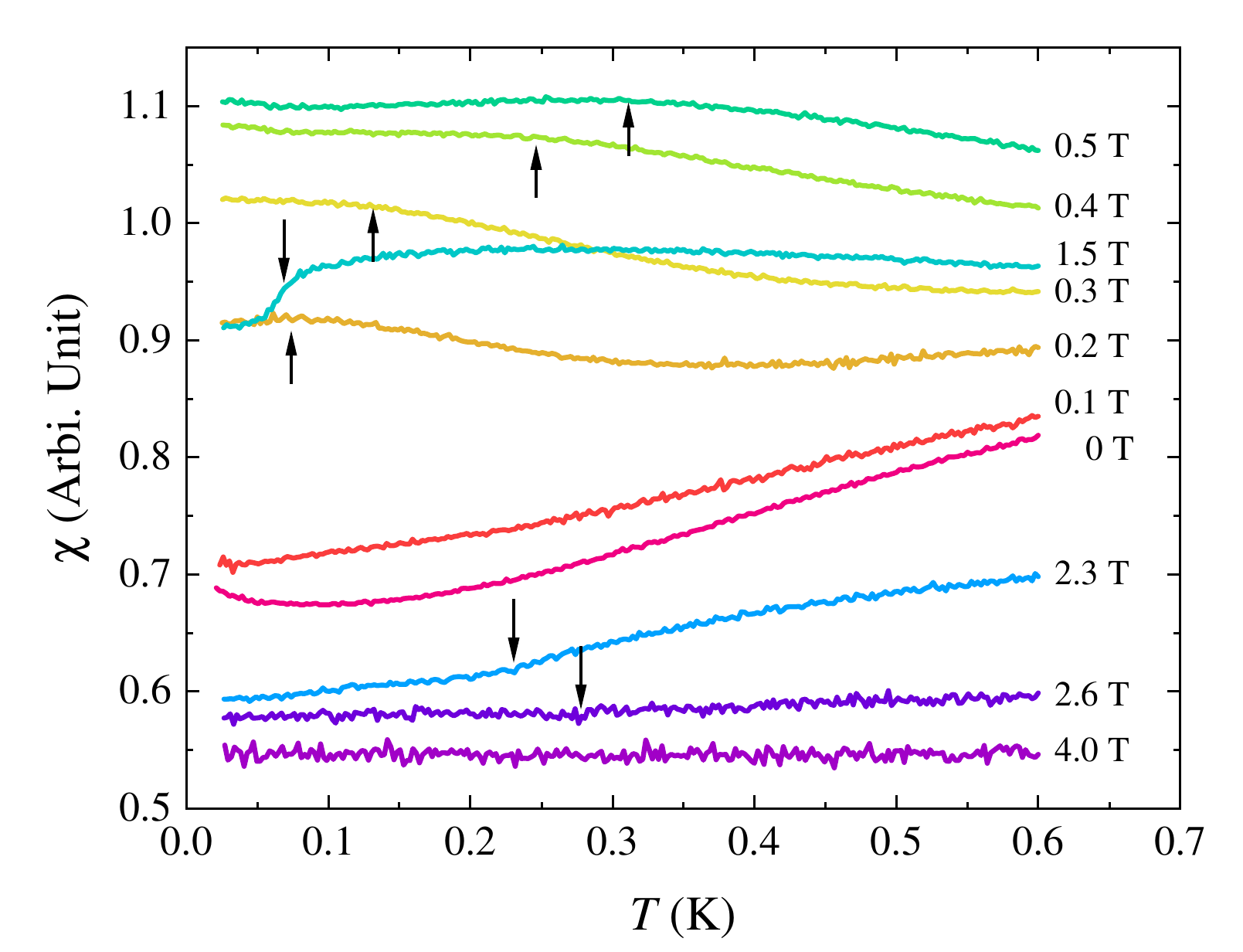}
    \caption{Temperature dependent ac susceptibility measured at various external dc fields. The black arrows indicate anomalies.}
    \label{ac}
\end{figure}
The heat capacity between 0.5 and 20 K shows no anomaly at zero magnetic field \cite{xu2024frustrated}. The ac magnetic susceptibility measured at zero dc field, shown in Fig.~\ref{ac}, likewise exhibits no anomalies down to 20 mK. These results suggest that the system does not develop magnetic long-range order, nor does a spin gap open at the lowest temperatures. However, the gradual decrease in the ac susceptibility, together with the dc susceptibility shown in Fig.~\ref{MvsT}, indicates the presence of a broad peak in the magnetic susceptibility between 0.6 and 2 K. The origin of this peak may be related to the development of spin–spin correlations below 3 K, as reported in Ref.~\cite{xu2024frustrated}. As the external dc magnetic field increases above 0.2 T, a broad hump whose peak location shifts towards high temperatures is observed. At 1.5 T, a sharp anomaly around 80 mK appears, which may be related to the weak 1/6 plateau. This peak also moves to higher temperature with increasing magnetic field. Above 4.0 T, no feature in ac susceptibility is observed between 20 mK and 0.6 K. 

\begin{figure}
    \centering
    \includegraphics[width=1\columnwidth]{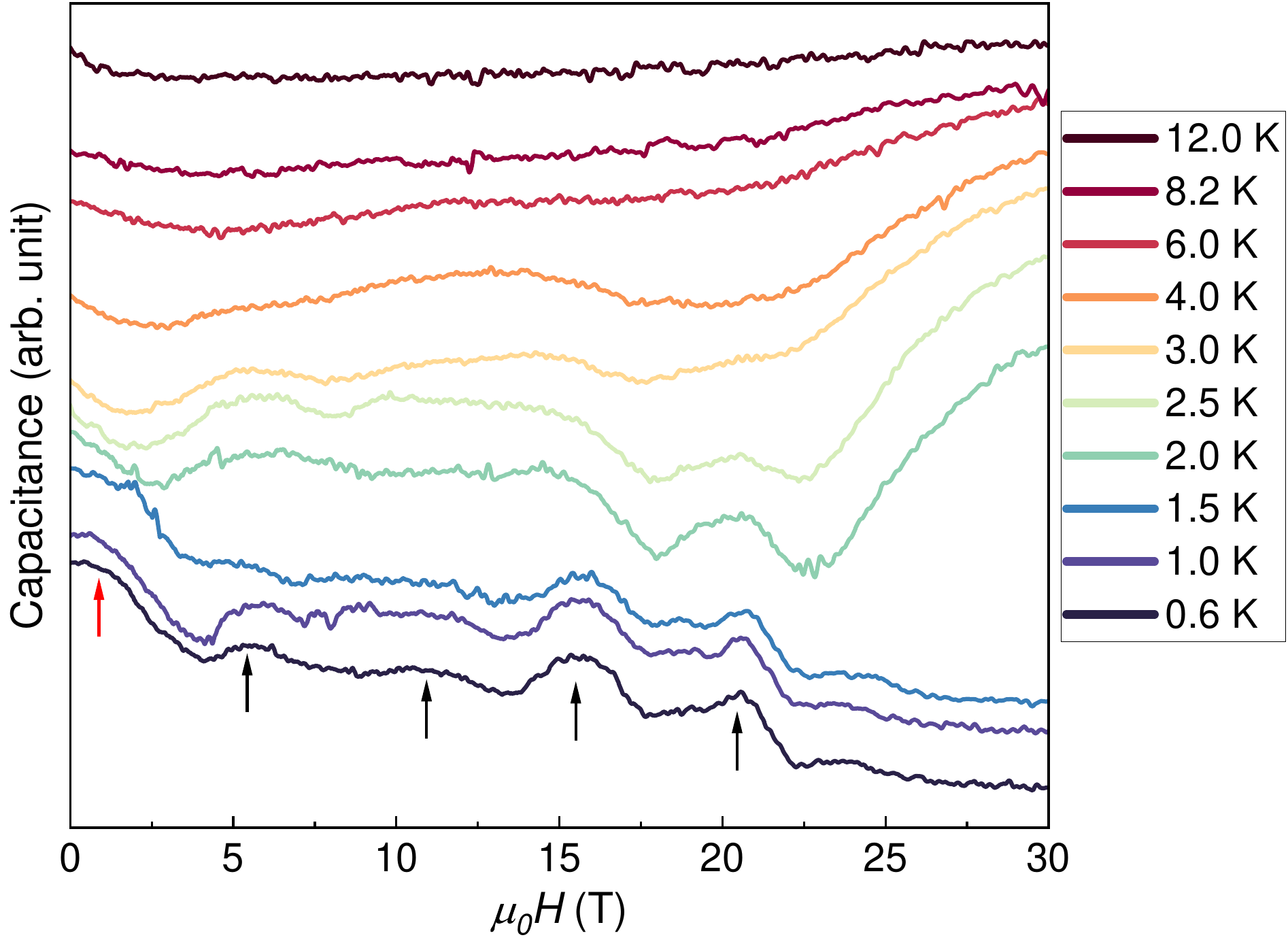}
    \caption{Capacitance as a function of pulsed magnetic field at various temperatures.}
    \label{Capacitance}
\end{figure}
Lastly, we measured the capacitance of the sample under magnetic field to examine whether the plateau phases are coupled to the electrical properties and/or accompanied by dimensional changes of the compound, as shown in Fig.~\ref{Capacitance}. Interestingly, the capacitance proves to be highly sensitive to the emergence of plateau phases. Although the magnetization at 2.5 K exhibits only weak step-like features, the capacitance measured at the same temperature clearly reveals corresponding anomalies associated with the plateaus. Additional measurements of electric polarization (not shown) did not detect any measurable polarization.

While investigating the microscopic origin of the capacitance variations lies beyond the scope of this work, the results imply that the formation of magnetic plateaus may involve either subtle lattice distortions or changes in electric polarizability, reflecting variations in the distribution of electron clouds. Such effects could, in turn, modify the exchange interactions as the magnetic state evolves. In the spin models discussed in the following section, we have not incorporated lattice or exchange-coupling variations; however, these effects may provide valuable insight for a more comprehensive understanding of this compound in future studies.


\section{Theory}
\label{sec:Theory} 

\subsection{Model}
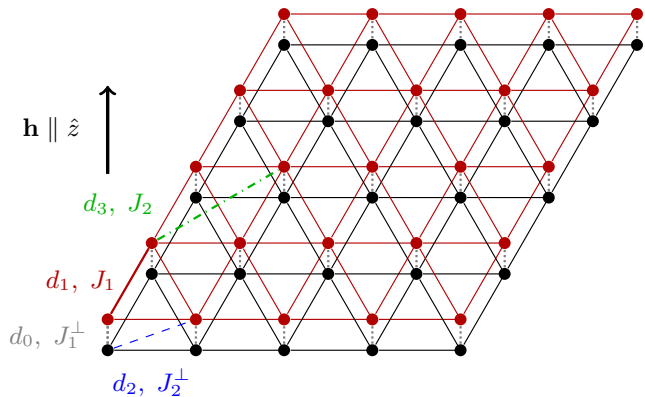
\begin{figure}[t]
\centering
\resizebox{\columnwidth}{!}{\begin{tikzpicture}[scale=1.1,
        every node/.style={circle, inner sep=1.5pt},
        layerA/.style={fill=black},      
        layerB/.style={fill=red!70!black}, 
        bond/.style={thin},
        inter/.style={densely dotted, gray, thick}
    ]

\def\nx{4}            
\def\ny{4}            
\def\dz{0.35}         
\def\rt{0.8660254}    

\newcommand{\Coord}[2]{\pgfmathparse{(#1)+0.5*(#2)}\let\x\pgfmathresult
                       \pgfmathparse{\rt*(#2)}\let\y\pgfmathresult}

\foreach \j in {0,...,\ny}{
  \foreach \i in {0,...,\nx}{
    \Coord{\i}{\j}
    \node[layerA] (A-\i-\j) at (\x,\y) {};
    \node[layerB] (B-\i-\j) at (\x,\y+\dz) {};
    
    \ifnum\i<\nx
      \Coord{\i+1}{\j}
      \draw[bond] (A-\i-\j) -- (\x,\y);
      \draw[bond,red!70!black] (B-\i-\j) -- (\x,\y+\dz);
    \fi
    
    \ifnum\j<\ny
      \Coord{\i}{\j+1}
      \draw[bond] (A-\i-\j) -- (\x,\y);
      \draw[bond,red!70!black] (B-\i-\j) -- (\x,\y+\dz);
    \fi
    
    \ifnum\i<\nx\relax\ifnum\j>0
      \Coord{\i+1}{\j-1}
      \draw[bond] (A-\i-\j) -- (\x,\y);
      \draw[bond,red!70!black] (B-\i-\j) -- (\x,\y+\dz);
    \fi\fi
    
    \draw[inter] (A-\i-\j) -- (B-\i-\j);
  }
}

\tikzset{cross/.style={dashed, blue}}

\draw[bond, thick, red!70!black] (B-0-0) -- (B-0-1)
      node[midway, left=2pt] {$d_1,\;J_1$};

\draw[bond, dash dot, thick, green!70!black] (B-0-1) -- (B-1-2)
      node[midway, left=20pt] {$d_3,\;J_2$};

\draw[inter] (A-0-0) -- (B-0-0)
      node[midway, left=4pt] {$d_0,\;J_1^{\perp}$};

\draw[cross] (A-0-0) -- (B-1-0)
      node[midway, below=0pt] {$d_2,\;J_2^{\perp}$};

\Coord{0}{0}                       
\coordinate (FieldStart) at (\x,\y+2);        

\pgfmathsetmacro\ArrowLen{1}

\draw[very thick,->] (FieldStart) -- ++(0,\ArrowLen)
      node[midway, left=20pt, anchor=center] {$\mathbf h \parallel \hat{z}$};

\end{tikzpicture}}
\caption{\textbf{Lattice structure of \(\mathrm{Rb_2Co_2(SeO_3)_3}\).} 
Magnetically active \(\mathrm{Co^{2+}}\) ions occupy the vertices; 
the interlayer spacing satisfies \(d_0\!\ll\!d_{1,2,3}\).}
\label{fig:lattice}
\end{figure}

Because the precise nature of the exchange interactions in \(\mathrm{Rb_2Co_2(SeO_3)_3}\) is not yet resolved, 
we adopt a minimal model that captures its dominant easy-axis anisotropy and layered geometry.  
A superexchange mechanism favors antiferromagnetic Heisenberg interactions, 
and the pronounced Ising-like susceptibility motivates an XXZ Hamiltonian 
\(H = H_{\parallel}+H_{\perp}+H_h\) with \(U(1)\) spin-rotation symmetry:
\begin{align}
H_\perp &= \sum_{ij} J_{ij}^{\perp}\!\left[S^z_{i0} S^z_{j1}
+ \Delta^\perp_{ij}\!\left(S^x_{i0} S^x_{j1} + S^y_{i0} S^y_{j1}\right)\!\right],\\
H_\parallel &= \sum_{ij}\sum_{l=0,1} J_{ij}\!\left[S^z_{il} S^z_{jl}
+ \Delta_{ij}\!\left(S^x_{il} S^x_{jl} + S^y_{il} S^y_{jl}\right)\!\right],\\
H_h &= -h\sum_{i,l} S_{il}^z,
\end{align}
where \(l=0,1\) labels the two layers and \(\vb{h}=g\mu_B\vb{B}=h\hat z\).
Because the closest magnetic ions lie directly above and below one another, 
the nearest-neighbor interlayer exchange \(J_1^{\perp}>0\) is expected to be the dominant energy scale (Fig.~\ref{fig:lattice}) \cite{li2025differences}.

\subsection{Triplon condensation}
\label{ssec:triplon}
In the strong-dimer limit, where \(J_1^{\perp}\) greatly exceeds all other exchanges, each interlayer bond forms a singlet
\begin{align*}
\ket{s}_i=\frac{1}{\sqrt2}
\big(\ket{\uparrow}_{i0}\ket{\downarrow}_{i1}-\ket{\downarrow}_{i0}\ket{\uparrow}_{i1}\big),
\end{align*}
and supports triplet excitations (``triplons'')
\begin{align*}
\ket{t_{+}}_{i} &= t_{+,i}^\dagger\ket{0}
= \ket{\uparrow}_{i0}\ket{\uparrow}_{i1},\\
\ket{t_{0}}_{i} &= t^\dagger_{0,i}\ket{0}
= \tfrac{1}{\sqrt{2}}\big(\ket{\uparrow}_{i0}\ket{\downarrow}_{i1}
+ \ket{\downarrow}_{i0}\ket{\uparrow}_{i1}\big),\\
\ket{t_{-}}_{i} &= t^\dagger_{-,i}\ket{0}
= \ket{\downarrow}_{i0}\ket{\downarrow}_{i1},
\end{align*}
created by \(t_{\mu,i}^\dagger\) acting on the vacuum \(\ket{0}\) \cite{bond-operator}.

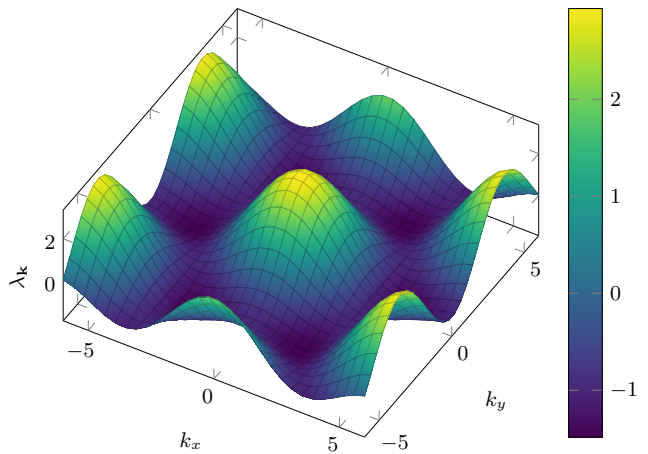
\begin{figure}[t]
\centering
\resizebox{\columnwidth}{!}{  \begin{tikzpicture}
    \begin{axis}[
        view={30}{80},
        width=\columnwidth,
        height=0.8\columnwidth,
        xlabel={$k_x$},
        ylabel={$k_y$},
        zlabel={$\lambda_{\vb{k}}$},
        domain=-6:6,
        y domain=-6:6,
        samples=30,
        colormap name=viridis,
        colorbar,
        z post scale=2.7,
    ]
    \pgfmathsetmacro{\ax}{1}
    \pgfmathsetmacro{\ay}{0}
    \pgfmathsetmacro{\bx}{0.5}
    \pgfmathsetmacro{\by}{0.8660254}

    \addplot3[surf, shader=faceted interp]{
        cos(deg(x*\ax + y*\ay)) +
        cos(deg(x*\bx + y*\by)) +
        cos(deg(x*(\bx - \ax) + y*(\by - \ay)))
    };
    \end{axis}
  \end{tikzpicture}}
\caption{\textbf{Normalized triplon dispersion \(\lambda_{\vb k}\).} 
Band minima occur at the \(K\) and \(K'\) points of the Brillouin zone.}
\label{fig:lambda}
\end{figure}

Intralayer couplings \(J_{ij}>0\) delocalize the triplons and hybridize their bands, producing a dispersion \(\lambda_{\vb k}\) with minima at the \(K\) and \(K'\) points of the Brillouin zone (Fig.~\ref{fig:lambda}).  
The lowest-energy branch corresponds to the \(t_0\) mode, decoupled from \(t_{\pm}\) by the \(U(1)\) symmetry.  
Increasing intralayer XXZ exchange induces Bose condensation of \(t_0\) triplons at \(K\) and \(K'\), which enlarges the magnetic unit cell and generates a three-sublattice long-range order in the zero-field ground state.

A magnetic field couples linearly to the triplon density,
\begin{align}
H_h &=-h\!\sum_{\vb k,s}\!
(t_{+,\vb k,s}^\dagger t_{+,\vb k,s}
-t_{-,\vb k,s}^\dagger t_{-,\vb k,s}),
\end{align}
where \(s=1,2,3\) labels the three sublattices of the enlarged unit cell.
The Zeeman term lowers the \(t_+\) branch while leaving the singlet and \(t_0\) sectors field-independent.
As \(h\) increases, the minimum of \(t_+\) band softens and condenses at the center of the magnetic Brillouin zone, triggering a quantum phase transition.  
Each magnetic unit cell then hosts one condensed \(t_+\) triplon and two mixed \((s,t_0)\) states, yielding a quantized magnetization \(m=1/3\) of the saturation value. 
The excitation gap of the uncondensed triplons stabilizes this plateau.

Within this triplon condensation picture, a finite triplon gap would produce an \(m=0\) magnetization plateau, yet experiments reveal a continuous magnetization curve with no such gap.  
This discrepancy indicates additional anisotropic exchange interactions beyond the XXZ model.  
Because the effective moments originate from spin-orbit coupled \(j_{\mathrm{eff}}=\tfrac12\) states of \(\mathrm{Co^{2+}}\), their residual orbital character naturally permits \(U(1)\)-symmetry-breaking couplings \(H'\)(such as Kitaev interactions or symmetric off-diagonal exchanges) that hybridize the \(S^z=0\) (\(s,t_0\)) and \(S^z=\pm1\) (\(t_\pm\)) sectors.
The ground state then becomes a mixed condensate containing finite \(t_\pm\) components in addition to the singlet and \(t_0\) triplon.
These \(S^z = \pm 1\) contributions couple directly to the  magnetic field, allowing the magnetization to grow smoothly even before the onset of a subsequent triplon condensation for the low-field plateaus.
Classically, \(t_\pm\) condensate permits continuous spin canting out of the plane under magnetic fields.
Therefore, the experimental absence of the triplon gap suggests the presence of $U(1)$ symmetry-breaking interactions that generate spin components aligned with the external magnetic field.

\subsection{Effective pseudospin model}
Our analysis so far has addressed the weak-field regime.  
We now turn to high magnetic fields and construct an effective theory that clarifies the origin of the observed magnetization plateaus.

At the saturation field \(h_s = J_1^{\perp}(1 + \Delta_{1}^{\perp})/2\), the energy levels of \(\ket{s}_i\) and \(\ket{t_{+}}_i\) cross, rendering the ground state twofold degenerate.  
Near \(h \!\approx\! h_s\), the low-energy Hilbert space on each dimer is therefore spanned by these two states,
\(
\{ \ket{s}_i, \ket{t_{+}}_i \} \equiv \{ \ket{0}_i, \ket{1}_i \}.
\)
Let us introduce a pseudospin-\(1/2\) operator \(\vb{\tilde{S}}_i\) acting on this subspace, defined by  
\(\tilde{S}^z_i \ket{n}_i = (-1)^n\tfrac12 \ket{n}_i\).  
For a system of \(N\) dimers, the saturated magnetization is \(M_s = N\).  
Defining the pseudospin magnetization \(\tilde{M} = \sum_i \tilde{S}^z_i\) with \(\tilde{M}_s = N/2\), the physical magnetization \(m = M / M_s\) and pseudospin magnetization \(\tilde{m} = \tilde{M} / \tilde{M}_s\) satisfy
\begin{align}
m = \tfrac{1}{2}(1 - \tilde{m}),
\label{eq: m_vs_tildem}
\end{align}
so that each plateau in \(m\) for the bilayer model corresponds to a plateau in \(\tilde{m}\) for the single-layer pseudospin model (Table~\ref{tab:m_vs_tildem}).

\begin{table}[t]
  \centering
  \renewcommand{\arraystretch}{1.5}
  \setlength{\tabcolsep}{9pt}
  \small
  \begin{tabular}{c|cccccc}
    \hline
    \(m\) & \(\frac16\) & \(\frac13\) & \(\frac12\) & \(\frac23\) & \(\frac56\) & \(1\) \\
    \hline
    \(\tilde m\) & \(\frac23\) & \(\frac13\) & \(0\) & \(-\frac13\) & \(-\frac23\) & \(-1\) \\
    \hline
  \end{tabular}
  \caption{Correspondence between physical-spin (\(m\)) and pseudospin (\(\tilde m\)) magnetization plateaus.}
  \label{tab:m_vs_tildem}
\end{table}

When the Hamiltonian \(H\) respects \(U(1)\) spin-rotation symmetry, the projected effective Hamiltonian \(\tilde H_{\text{eff}} = P H P\) retains the same symmetry and takes the canonical XXZ form:
\begin{multline}
\tilde H_{\text{eff}}
= - \sum_{i}\tilde h_{i}\tilde S^{z}_{i}
\\
+\sum_{ij}\tilde J_{ij}\!\left[
\tilde S^{z}_{i}\tilde S^{z}_{j}
+\tilde\Delta_{ij}\!\left(\tilde S^{x}_{i}\tilde S^{x}_{j} + \tilde S^{y}_{i}\tilde S^{y}_{j}\right)
\right],
\label{eq:Heff}
\end{multline}
where
\begin{align*}
\tilde J_{ij} &= \tfrac{1}{4}\!\left(J_{ij}+J^{\perp}_{ij}\right),\\
\tilde\Delta_{ij} &= 
\frac{2\!\left(J_{ij}\Delta_{ij}-J^{\perp}_{ij}\Delta^{\perp}_{ij}\right)}
{J_{ij}+J^{\perp}_{ij}},\\
\tilde h_{i} &= -h + \tfrac{1}{4}\!\sum_{j}\!\left(J_{ij}+J^{\perp}_{ij}\right).
\end{align*}

With the original Hamiltonian, quantum nature of the singlet states make large-scale calculations intractable, while geometric frustration on the triangular lattice introduces a sign problem that rules out quantum Monte Carlo simulations.  
In the pseudospin formulation each dimer reduces to a local two-level system.  
Although frustration still hinders a full quantum treatment, valuable insight can be gained from classical simulations in the Ising limit \(\tilde{\Delta}_{ij}\!\to\!0\), where quantum fluctuations are frozen.

For a uniform longitudinal field \(\tilde h < 0\), the familiar ``down–down–up'' (triplet–triplet–singlet) configuration on the triangular lattice stabilizes the plateau at \(\tilde m=-\tfrac13\) (\(m=\tfrac23\)).  
The fully polarized state \(\tilde m=-1\) (\(m=1\)) appears when the field is large enough to align all pseudospins.  
Including a next-nearest-neighbor Ising coupling \(\tilde J_2\) favors a stripe phase \cite{J2Ising}, producing a plateau at \(\tilde m=0\) (\(m=\tfrac12\)).  
Further-neighbor interactions, particularly a fourth-neighbor term \(\tilde J_4\), are known to stabilize the plateau at \(\tilde m=-\tfrac23\) (\(m=\tfrac56\)) \cite{J4Ising}.  
As a proof of principle, the extended Ising pseudospin model reproduces the family of plateaus at \(m=\tfrac12,\tfrac23,\tfrac56, 1\) observed experimentally in \(\mathrm{Rb_2Co_2(SeO_3)_3}\).  
The extended Ising pseudospin model also yields additional plateaus at \(\tilde m=-\tfrac12\) (\(m=\tfrac34\)) and \(\tilde m=-\tfrac79\) (\(m=\tfrac89\)), but they may be destabilized by quantum fluctuations or competing interactions in more realistic models.

\section{Discussion}
Recently, there has been an intense discussion regarding the magnetic properties of \rcs{} and K$_{2}$Co$_{2}$(SeO$_{3}$)$_{3}$. The main debate centers on whether the system should be understood from a spin-dimer perspective, assuming dominant antiferromagnetic Heisenberg exchange interactions between nearest-neighbor Co ions, or from a nearly free-spin viewpoint, in which the Co–O–Co bond angle close to 90$^{\circ}$ induces ferromagnetic exchange interactions that partially cancel the antiferromagnetic ones, resulting in a very weak net coupling between neighboring Co spins. Models constructed based on each of these approaches were able to account only for the 1/3 magnetization plateau observed at low magnetic fields.

In our analysis, we began from a coupled dimer model. This choice was motivated by two key observations: first, inelastic neutron scattering results indicate that the magnetic excitations cannot be explained solely by Heisenberg exchange interactions; and second, the presence of pure Heisenberg exchange would preserve U(1) spin-rotational symmetry, which is inconsistent with the finite slopes observed in the magnetization plateaus. Therefore, we adopted a more generalized spin Hamiltonian. Remarkably, our model successfully reproduces all of the experimentally observed plateaus, both in the low- and high-field regions, for the first time. Interestingly, the analysis highlights the crucial role of the bond-dependent anisotropy term. The nearly negligible spin gap at zero field further suggests that this bond-dependent anisotropy may be substantially larger than previously assumed.

The first possible form of bond-dependent anisotropy to consider is the Dzyaloshinskii–Moriya (DM) interaction. As shown in the Appendix, although the DM interaction is symmetry-allowed in principle, the three Co–O–Co bonds related by $C_{3}$ rotation in the structure possess orientations such that their respective DM vectors nearly cancel each other, resulting in an almost negligible net DM interaction. 

The next candidates for anisotropic exchange are the Kitaev-type and $\Gamma$-type interactions. In cobalt-based materials, interest in these bond-dependent terms has recently intensified, driven by the question of whether Kitaev physics can be realized in Co-based compounds. In \rcs{} the two CoO$_{6}$ octahedra share a face, and the Co–O–Co bond angle is slightly distorted from the ideal 90$^{\circ}$. Under such geometric conditions, certain bond-dependent anisotropy terms could remain finite and influence the magnetic properties; however, their microscopic origin and magnitude have not yet been quantitatively established. Our experimental results and modeling strongly underscore the need for detailed microscopic calculations to clarify the role of such anisotropic exchanges in this compound.

The existence of the 1/6 plateau has also been a subject of considerable discussion. In the field region where the 1/6 plateau is expected, a subtle change in slope is observed in the magnetization curve—so subtle that it becomes apparent only after taking the derivative. Even at the lowest temperature investigated, 20 mK, which is the lowest among all studies to date, the feature does not sharpen into a distinct plateau. Although further investigation is needed to clarify the nature of this phase, the accompanying anomaly in capacitance near the 1/6 plateau suggests that it may correspond to a crossover-like transition involving a very small change in electric polarizability.

\section{Conclusion}
In this study, we investigated the high-field magnetization properties of \rcs{}. We observed four or five distinct magnetization plateaus, and the absence of an energy gap at zero magnetic field, together with the finite slopes of all plateaus, indicates that the U(1) spin-rotational symmetry is not preserved. From the capacitance measurements, we further infer that each plateau is accompanied by changes in either the lattice dimensions or the electric polarizability—both of which can modify the exchange interactions in the system. We also have developed a theoretical framework that accounts for the sequence of observed magnetization plateaus.
Assuming dominant antiferromagnetic interlayer coupling, successive triplon condensations naturally explain the low-field plateaus, while an effective pseudospin model captures the high-field regime near saturation.  
Although the full microscopic spin Hamiltonian remains unknown, these complementary descriptions---interlayer dimerization, triplon condensation, and dual pseudospin theory---offer a unified picture consistent with experiment.  

\begin{acknowledgments}
We thank Hao Zhang and Cristian Batista for the useful discussion. The experimental work (S.Z., G.S.F., V.S.Z, and M.L.) at LANL was primarily funded by the LDRD program at LANL. The facilities of the NHMFL are funded by the U.S. NSF through Cooperative Grant No. DMR-1644779, the U.S. DOE, and the State of Florida.
The theoretical work at LANL (S-Z.L.) was carried out under the auspices of the U.S. DOE NNSA under contract No. 89233218CNA000001 through the LDRD Program, and was supported by the Center for Nonlinear Studies at LANL (W.C.), and was performed, in part, at the Center for Integrated Nanotechnologies, an Office of Science User Facility operated for the U.S. DOE Office of Science, under user proposals $\#2018BU0010$ and $\#2018BU0083$.
\end{acknowledgments}

\newpage
\appendix

\section{Bond operator representations of correlated bilayers}
In Sec.~\ref{ssec:triplon}, we discussed the four local basis states \(\{ \ket {s}_i, \ket{t_+}_i, \ket{t_0}_i, \ket{t_-}_i\}\) in the strong interlayer coupling limit.
Because the four local states form a complete basis on each dimer at site \(i\), we employ the bond‑operator representation introduced by Sachdev and Bhatt~\cite{bond-operator}.

With the bosonic creation and annihilation operators of the singlet and triplet states satisfying \([s_i, s_j^\dagger] = \delta_{ij}\), \([t_{\mu,i}, t^\dagger_{\nu,j}] = \delta_{\mu\nu}\delta_{ij}\), at each in-plane position \(i, j\), the spin operators can be expressed as follows:
\begin{align*}
&S^{x}_{i,0}S^{x}_{i,1} + S^{y}_{i,0}S^{y}_{i,1}
= \frac{1}{2}\left( t_{0,i}^\dagger t_{0,i} - s_i^\dagger s_i \right),
\\
&S^{z}_{i,0}S^{z}_{i,1}
= \frac{1}{4}\left(t_{+,i}^\dagger t_{+,i} -t_{0,i}^\dagger t_{0,i} + t_{-,i}^\dagger t_{-,i} - s_i^\dagger s_i \right),
\\
&S^{+}_{i,0}
= \frac{1}{\sqrt{2}}\left(t_{+,i}^\dagger \left(t_{0,i} - s_i\right) + \left(t_{0,i}^\dagger + s_i^\dagger \right) t_{-,i} \right)
=\left(S^{-}_{i,0} \right)^\dagger,
\\
&S^{z}_{i,0}
= \frac{1}{2} \left(t_{+,i}^\dagger t_{+,i} - t_{-,i}^\dagger t_{-,i} + s_i^\dagger t_{0,i} + t_{0,i}^\dagger s_i \right),
\\
&S^{+}_{i,1}
= \frac{1}{\sqrt{2}}\left(t_{+,i}^\dagger \left(t_{0,i} + s_{i} \right) + \left(t_{0,i}^\dagger-s_i^\dagger\right) t_{-,i} \right)
=\left( S^{-}_{i,1}\right)^\dagger,
\\
&S^{z}_{i,1}
=\frac{1}{2}\left( t_{+,i}^\dagger t_{+,i} - t_{-,i}^\dagger t_{-,i} - s_i^\dagger t_{0,i} - t_{0,i}^\dagger s_i\right),
\end{align*}
where \(S^{\pm} = S^{x} \pm i S^{y}\).
The nearest-neighbor interlayer XXZ interaction is then
\begin{align}
H_{0} &= \sum_{i} J_{1}^{\perp} \left[
S^z_{i,0} S^z_{i,1}
+\Delta_1^{\perp}
\left(
S_{i,0}^{x}S_{i,1}^{x} + S_{i,0}^{y}S_{i,1}^{y}
\right)
\right]
\nonumber \\
&=\sum_{i}\frac{J_{1}^{\perp}}{4}
\Big[
t^{\dagger}_{+,i}t_{+,i}+t^{\dagger}_{-,i}t_{-,i}
\nonumber \\
&\qquad\quad
-\left(1-2\Delta_{1}^{\perp}\right)t^{\dagger}_{0,i}t_{0,i}
-\left(1+2\Delta_{1}^{\perp}\right)s^{\dagger}_{i}s_{i}
\Big].
\end{align}
The spectrum of \(H_{0}\) is bounded from below upon imposing the single‑occupancy constraint that enforces \(S=1/2\) on every site,
\begin{align}
\left(\vb{S}_{il}\right)^2
&=\frac{3}{4}\left( t_{+,i}^\dagger t_{+,i} + t_{0,i}^\dagger t_{0,i} + t_{-,i}^\dagger t_{-,i} + s_i^\dagger s_i \right)
\nonumber \\
&= S(S+1) = \frac{3}{4}
\nonumber \\
\Rightarrow  t_{+,i}^\dagger & t_{+,i} + t_{0,i}^\dagger t_{0,i} + t_{-,i}^\dagger t_{-,i} + s_i^\dagger s_i = 1.
\end{align}

Together, the bond‑operator representation and this local constraint provide a concrete starting point for analyzing inter‑dimer interactions and the quantum phase transition triggered by triplon condensation.

In the bond-operator language, the ground state of the dominant interlayer Hamiltonian \(H_0\) is the fully condensed singlet state such that \(\langle s_i \rangle = 1\).
When additional interactions and an external magnetic field are introduced, triplet excitations become progressively more relevant.
Provided that the interlayer coupling \(J_1^{\perp}\) remains dominant, the singlet occupation on each site is expected to remain significant.
Therefore, we approximate the singlet operators by their nonzero expectation values, \(s_i, s_i^\dagger \to \langle s_i \rangle = s \neq 0\), whose precise value is determined self-consistently.

To illustrate this explicitly, consider the nearest-neighbor intralayer XXZ interaction.
Using the bond-operator representation and adopting the mean-field approximation for singlets, the nearest-neighbor XXZ coupling takes the form:
\begin{widetext}
\begin{align}
H_1 &= J_1 \sum_{\langle i, j\rangle} \sum_{l=0,1} S_{i,l}^z S_{j,l}^z
+ \frac{\Delta_1}{2} \left( S^{+}_{i,l}S^{-}_{j,l} + S^{-}_{i,l}S^{+}_{j,l} \right)
\\
&\approx \frac{J_1}{2} \sum_{\langle i,j \rangle} s^2
\left( t_{0,i}^\dagger t_{0,j} + t_{0,j}^\dagger t_{0,i} + t_{0,i}t_{0,j} + t_{0,i}^\dagger t_{0,j}^\dagger \right)
\nonumber \\
&\qquad\qquad
+\frac{1}{2} \left(
t_{+,i}^\dagger t_{+,i}t_{+,j}^\dagger t_{+,j} - t_{+,i}^\dagger t_{+,i}t_{-,j}^\dagger t_{-,j}
- t_{-,i}^\dagger t_{-,i} t_{+,j}^\dagger t_{+,j} + t_{-,i}^\dagger t_{-,i}t_{-,j}^\dagger t_{-,j}
\right)
\label{eq: H1zz} \\
&+\frac{\Delta_1}{2}\Bigg[
s^2 \Big(
t_{+,i}^\dagger t_{+,j} + t_{+,j}^\dagger t_{+,i} + t_{-,i}^\dagger t_{-,j} + t_{-,j}^\dagger t_{-,i}
- t_{+,i}^\dagger t_{-,j}^\dagger -t_{+,i}t_{-,j} - t_{-,i}^\dagger t_{+,j}^\dagger - t_{-,i}t_{+,j}
\Big)
\nonumber \\
&\qquad\quad
+\Big(
t_{+,i}^\dagger t_{0,i} t_{0,j}^\dagger t_{+,j} + t_{0,i}^\dagger t_{+,i} t_{+,j}^\dagger t_{0,j}
+ t_{+,i}^\dagger t_{0,i} t_{-,j}^\dagger t_{0,j} + t_{0,i}^\dagger t_{+,i} t_{0,j}^\dagger t_{-,j}
\nonumber \\
&\qquad\qquad\quad
+ t_{0,i}^\dagger t_{-,i} t_{0,j}^\dagger t_{+,j} + t_{-,i}^\dagger t_{0,i} t_{+,j}^\dagger t_{0,j}
+ t_{0,i}^\dagger t_{-,i} t_{-,j}^\dagger t_{0,j} + t_{-,i}^\dagger t_{0,i}t_{0,j}^\dagger t_{-,j}
\Big)
\Bigg].
\label{eq: H1xy}
\end{align}
\end{widetext}

When treating the Hamiltonian \(H = H_0 + H_1\), the local constraint of one boson per site must also be enforced.
Expressing this constraint through a delta function and introducing a gauge field \(a_j(t)\), we have
\begin{widetext}
\begin{align*}
\prod_j \delta\left(s_j^\dagger(t) s_j(t) + \sum_\mu t_{\mu,j}^\dagger(t) t_{\mu,j}(t) - 1\right)
\propto \int \mathcal{D}[a(t)] \, \exp\left[
\sum_j i a_j(t) \left(s_j^\dagger(t) s_j(t) + \sum_\mu t_{\mu,j}^\dagger(t) t_{\mu,j}(t) - 1\right)
\right].
\end{align*}
\end{widetext}
Hence, the full quantum partition function is
\begin{align*}
Z = \int \mathcal{D}[s^\dagger, s]\mathcal{D}[t_{\mu}^\dagger, t_{\mu}] \mathcal{D}[a] \, e^{i S[s^\dagger,s, t_{\mu}^\dagger, t_{\mu}, a]}
\end{align*}
with the action
\begin{align*}
S = \int \! dt \, \sum_j s_j^\dagger (i\partial_t - a_j) s_j + t_{\mu,j}^\dagger (i\partial_t - a_j) t_{\mu,j}- H.
\end{align*}

To tackle this interacting problem, we employ two simplifying approximations: (i) quantum fluctuations of the gauge field $a_j(t)$ are neglected, imposing the one-particle-per-site constraint only on average, and (ii) the quartic interactions are approximated through mean-field decoupling schemes that preserve the \(U(1)\) spin-rotation symmetry, thereby excluding spontaneous \(U(1)\) symmetry breaking.
Furthermore, since the ground state of \(H_0\) is nondegenerate and gapped, weak intralayer interactions \(H_1\) are not expected to spontaneously break other discrete crystalline symmetries.
Hence, we assume that translational invariance, sixfold lattice rotation, and inversion symmetry with respect to the bond centers remain intact.
Under these approximations, the mean-field Hamiltonian in momentum space becomes
\begin{align}
H_0 &= -\frac{J_1^{\perp}}{4}N\left(1+2 \Delta_1^{\perp}\right) s^2
+ \frac{J_1^{\perp}}{4} \sum_{\vb k} t_{+,\vb k}^\dagger t_{+, \vb k} + t_{-,\vb k}^\dagger t_{-, \vb k}
\nonumber \\
&- \left(1-2\Delta_1^{\perp}\right) t_{0,\vb k}^\dagger t_{0, \vb k}, 
\label{eq: H0k}
\end{align}
and
\begin{align}
H_1 &= \frac{J_1}{4} \sum_{\vb k} \sum_{\mu,\nu} \lambda_{\vb k} \varphi_{\mu}^\dagger(\vb k) \mathcal{M}_{\mu\nu}\varphi_{\nu}(\vb k),
\end{align}
with
\(
\varphi(\vb k) = 
\left(
t_{+,\vb k},\, t^\dagger_{-,-\vb k},\, t_{-,\vb k},\, t^\dagger_{+,-\vb k},\, t_{0,\vb k},\, t^\dagger_{0,-\vb k}
\right)^T,
\)
and the block-diagonal matrix
\begin{align}
\mathcal{M}=
\begin{pmatrix}
\mathcal{M}_{+-} & 0 & 0 \\
0 & \mathcal{M}_{-+} & 0 \\
0 & 0 & 2\mathcal{M}_0
\end{pmatrix}.
\label{eq: M}
\end{align}
The momentum dependence arises solely from
\begin{align}
\lambda_{\vb k} = \cos (\vb k\! \cdot \!\vb a_1) + \cos (\vb k\! \cdot \!\vb a_2 ) 
+ \cos (\vb k \!\cdot \!(\vb a_2 - \vb a_1)),
\end{align}
with primitive lattice vectors $\vb a_1 = (1,0)$, $\vb a_2 = \tfrac{1}{2}(1,\sqrt3)$.
The block matrices are explicitly
\begin{widetext}
\begin{align*}
\mathcal{M}_{+-} &=
\begin{pmatrix}
3(n^+ - n^-) + \chi^{+} +\Delta_1 \left(\chi^0 + s^2\right) &  -\eta^{\pm} + \Delta_1 (\eta^0 - s^2)  \\
-(\eta^{\pm})^* + \Delta_1 \big((\eta^0)^* - s^2\big) & -3(n^+ - n^-) + \chi^{-} + \Delta_1 \left(\chi^0 + s^2\right)
\end{pmatrix},
\\
\mathcal{M}_{-+} &=
\begin{pmatrix}
-3(n^+ - n^-) + \chi^{-} +\Delta_1\left(\chi^0 + s^2\right) &  -\eta^{\pm} + \Delta_1 \left(\eta^0 - s^2\right)  \\
-(\eta^{\pm})^* + \Delta_1 \left((\eta^0)^* - s^2\right) & 3(n^+ - n^-) + \chi^{+} + \Delta_1 \left(\chi^0 + s^2\right)
\end{pmatrix},
\\
\mathcal{M}_0 &=
\begin{pmatrix}
\tfrac{1}{2}\left( \chi^{+} + \chi^{-} \right) + s^2 & \eta^{\pm} + s^2 \\
\left(\eta^{\pm} \right)^* + s^2 & \tfrac{1}{2}\left( \chi^{+} + \chi^{-} \right) + s^2 
\end{pmatrix},
\end{align*}
\end{widetext}
with the mean-field parameters defined as
\begin{align}
n^{\mu} &= \langle t_{\mu,i}^\dagger t_{\mu,i} \rangle,
\quad
\chi^{\mu} = \langle t_{\mu,i}^\dagger t_{\mu,j} \rangle,
\\
\eta^{\pm} &= \langle t_{+,i} t_{-,j} \rangle = \langle t_{-,i} t_{+,j} \rangle,
\quad
\eta^0 = \langle t_{0,i} t_{0,j} \rangle,
\end{align}
for the nearest-neighbor sites \(i, j\).
The block-diagonal structure of \(\mathcal M\) reflects the \(U(1)\) spin rotation symmetry.

While the self-consistent mean-field parameters determine the detailed structure of the eigenvectors, the condensation momentum \(\vb k_0\) can be obtained by minimizing \(\lambda_{\vb k}\) without knowing the self-consistent solution.
Since \(\lambda_{\vb k}\) attains its minima at the corners (\(K\) and \(K'\) points) of the first Brillouin zone, the condensation momentum is \(\vb k_0 = K, K'\).

\section{Derivation of the pseudospin model}

Consider a single dimer at in-plane position \(i\) in a strong magnetic field \(h\):
\begin{multline*}
H_{\text{dimer}} = J_1^{\perp}\big[ S^z_{i,0} S^z_{i,1}
+ \Delta^{\perp}_1 \left( S^x_{i,0} S^x_{i,1} + S^y_{i,0} S^y_{i,1} \right)\big]
\\ 
- h \left(S^z_{i,0} + S^z_{i,1} \right).
\end{multline*}
The eigenstates are still singlets and triplets, but the eigenvalues of the triplets \(\ket{t_{\pm}}_i\) acquire Zeeman shifts:
\begin{align*}
\ket{s}_i, \quad &\varepsilon_s = -\frac{J_1^{\perp}}{4}\left( 1 + 2\Delta_1^{\perp}\right),
\\
\ket{t_{+}}_i, \quad &\varepsilon_{+} = \frac{J_1^{\perp}}{4} - h,
\\
\ket{t_0}_i, \quad &\varepsilon_0 = -\frac{J_1^{\perp}}{4}\left( 1 - 2\Delta_1^{\perp}\right),
\\
\ket{t_{-}}_i, \quad &\varepsilon_{-} = \frac{J_1^{\perp}}{4} + h.
\end{align*}
At the saturation field \(h_s= \frac{J_1^{\perp}}{2}\left(1 + \Delta_{1}^{\perp} \right)\), the energy levels for \(\ket{s}_i\) and \(\ket{t_{+}}_i\) cross, making the ground state exactly twofold degenerate.
Near \(h\approx h_s\), the low-energy Hilbert space on each dimer is therefore spanned by these two states, which we label
\begin{align}
\ket{0}_i \equiv \ket{s}_i,\qquad \ket{1}_i \equiv \ket{t_{+}}_i.
\end{align}

Let us introduce a pseudospin-1/2 operator $\vb{\tilde{S}}_i$ acting on this subspace by
\begin{align}
    \tilde{S}^z_i \ket{0}_i = + \frac{1}{2}\ket{0}_i, \qquad \tilde{S}^z_i \ket{1}_i = - \frac{1}{2}\ket{1}_i
\end{align}
and define the projector onto the pseudospin subspace,
\begin{align}
P = \prod_i \sum_{n=0,1} \ket{n}_i \bra{n}_i.
\end{align}
Since $\{\ket{0}_i, \ket{1}_i\}$ are also eigenstates of the physical spin operators $S^z_i \equiv S^z_{i,0} + S^z_{i,1}$ such that
\begin{align}
S^z_i \ket{0}_i = 0\ket{0}_i, \qquad  S^z_i \ket{1}_i = 1\ket{1}_i,
\end{align}
a short calculation gives
\begin{align}
P S_i^z P = \frac{1}{2} - \tilde{S}^z_i.
\end{align}

When \(|h-h_s|\) is small compared with other energy scales, this pseudospin description remains a good approximation.
The effective theory developed below therefore targets the large-\(m\) plateaus under high magnetic fields.

Projecting the intralayer and further-neighbor interlayer interactions onto the low-energy pseudospin manifold defined by the projector \(P\).
Each spin operator reduces to
\begin{align*}
P S^{z}_{i,0} P &= P S^{z}_{i,1} P \;=\; \frac14\left(1-2\tilde S^{z}_{i}\right),
\\
P S^{+}_{i,0} P &= \left(P S^{-}_{i,0} P\right)^{\dagger}
= -\frac{1}{\sqrt{2}}\left(\tilde S^{x}_{i}-i\tilde S^{y}_{i}\right)
= -\frac{1}{\sqrt{2}}\,\tilde S^{-}_{i},
\\
P S^{+}_{i,1} P &= \left(P S^{-}_{i,1} P\right)^{\dagger}
= +\frac{1}{\sqrt{2}}\left(\tilde S^{x}_{i}-i\tilde S^{y}_{i}\right)
= +\frac{1}{\sqrt{2}}\,\tilde S^{-}_{i}.
\end{align*}
Using these identities, we obtain
\begin{align*}
&P S^{z}_{i,l} S^{z}_{j,l'} P
= \frac{1}{16} - \frac{1}{8}\left(\tilde S^{z}_{i}+\tilde S^{z}_{j}\right)
+ \frac{1}{4}\,\tilde S^{z}_{i}\tilde S^{z}_{j},
\\
&P\!\left(S^{x}_{i,l} S^{x}_{j,l} + S^{y}_{i,l} S^{y}_{j,l}\right)\!P
= \frac{1}{2}\left(\tilde S^{x}_{i}\tilde S^{x}_{j}+\tilde S^{y}_{i}\tilde S^{y}_{j}\right), 
\\
&P\!\left(S^{x}_{i,l} S^{x}_{j,1-l}+ S^{y}_{i,l} S^{y}_{j,1-l}\right)\!P
= -\frac{1}{2}\left(\tilde S^{x}_{i}\tilde S^{x}_{j}+\tilde S^{y}_{i}\tilde S^{y}_{j}\right).
\end{align*}

The projected intralayer and interlayer XXZ interactions and the Zeeman coupling are therefore
\begin{widetext}
\begin{align}
P H_{\parallel} P
&= \sum_{i,j}\sum_{l=0,1} J_{ij}\,
P\!\left[ S^{z}_{i,l}S^{z}_{j,l}
+ \Delta_{ij}\left(
S^{x}_{i,l}S^{x}_{j,l}+S^{y}_{i,l}S^{y}_{j,l}
\right)\right]\!P
\nonumber \\
&= \frac{1}{16}\sum_{i,j}J_{ij}
-\frac{1}{4} \sum_{i}\Bigl(\sum_{j}J_{ij}\Bigr)\tilde S^{z}_{i}
+\frac{1}{4} \sum_{i,j}J_{ij}\left[
\tilde S^{z}_{i}\tilde S^{z}_{j}
+ 2\Delta_{ij}\left(\tilde S^{x}_{i}\tilde S^{x}_{j}+\tilde S^{y}_{i}\tilde S^{y}_{j}\right)
\right],
\\
P H_{\perp} P
&= \sum_{i,j} J^{\perp}_{ij}\,
P\!\left[ S^{z}_{i,0}S^{z}_{j,1}
+ \Delta^{\perp}_{ij}\left(S^{x}_{i,0}S^{x}_{j,1}+S^{y}_{i,0}S^{y}_{j,1}\right)
\right]\!P 
\nonumber \\
&= \frac{1}{16}\sum_{i,j}J^{\perp}_{ij}
-\frac{1}{4}\sum_{i}\Bigl(\sum_{j}J^{\perp}_{ij}\Bigr)\tilde S^{z}_{i}
+\frac{1}{4} \sum_{i,j}J^{\perp}_{ij}\left[
\tilde S^{z}_{i}\tilde S^{z}_{j}
- 2\Delta^{\perp}_{ij}\left(\tilde S^{x}_{i}\tilde S^{x}_{j}
+\tilde S^{y}_{i}\tilde S^{y}_{j}\right)
\right],
\\
PH_{h}P&=-h\sum_{i,l}PS^{z}_{i,l}P = h\sum_i \tilde S_i^z + \text{const.}
\end{align}
\end{widetext}
Adding these terms yields the effective pseudospin Hamiltonian in Eq.~(\ref{eq:Heff}).

When the original spin Hamiltonian respects \(U(1)\) spin-rotation symmetry about the \(z\)-axis, i.e., \([H,M]=0\) with \(M=\sum_{i,l}S^{z}_{i,l}\), the projection \(P\) inherits this symmetry:
\begin{align}
[\tilde H_{\text{eff}},\tilde M] =[P H P,\,P M P] =P\,[H,M]\,P = 0,
\end{align}
because \([M, P] = 0\) and \(P M P = N/2 - \tilde{M}\).
Hence, \(\tilde H_{\text{eff}}\) preserves the \(U(1)\) rotation symmetry of the pseudospins about their \(z\)-axis.
Therefore, Eq.~(\ref{eq:Heff}) constitutes the most general bilinear pseudospin Hamiltonian consistent with the symmetries of the original model in the strong-field limit.

\section{Symmetrically prohibited Dzyaloshinskii–Moriya interaction}
\begin{figure}
    \centering
    \includegraphics[width=1\columnwidth]{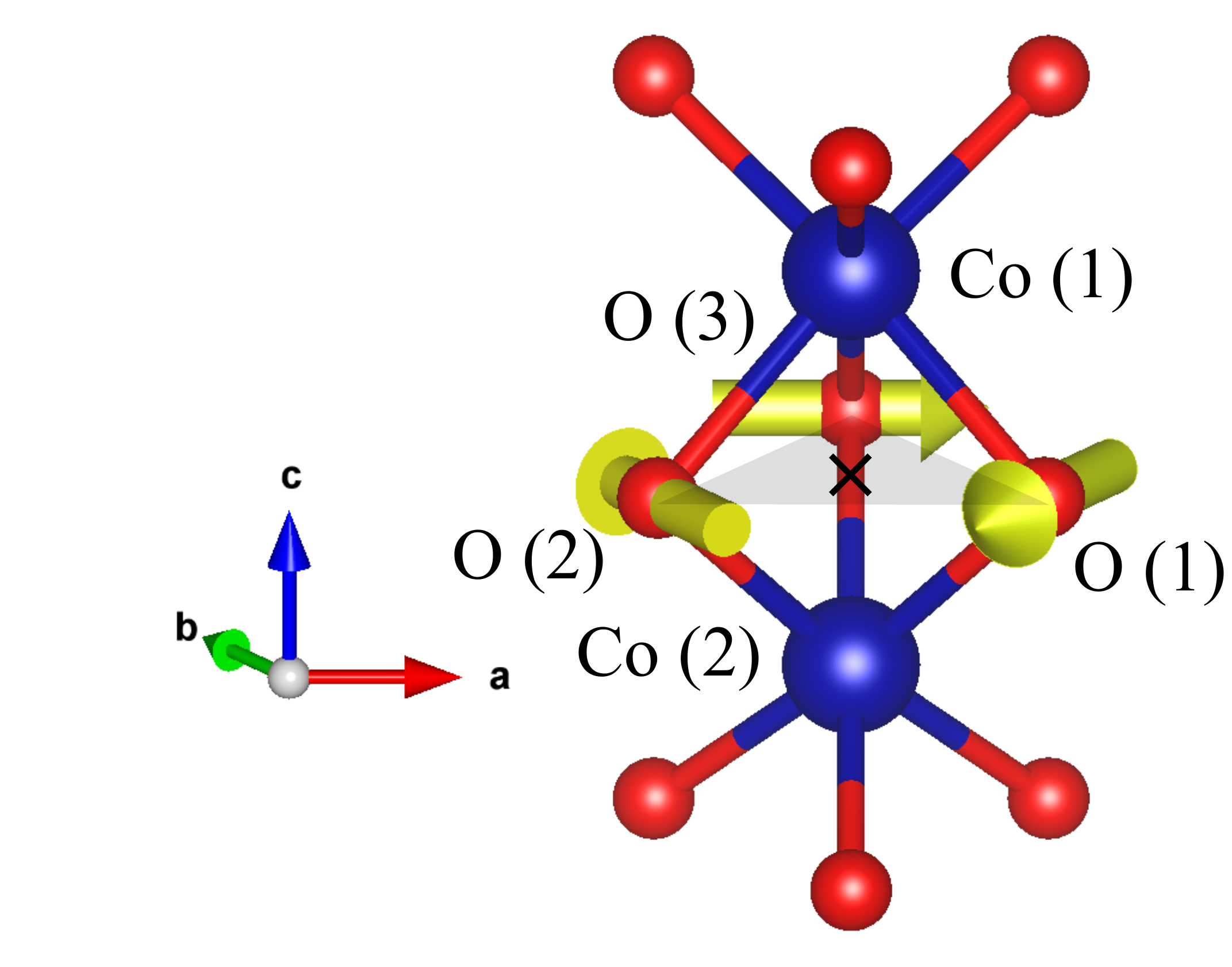}
    \caption{Local structure of a Co dimer. Red: Oxygen, Blue: Cobalts, Yellow arrows: Directions of DM interaction.}
    \label{DM}
\end{figure}
Fig.~\ref{DM} shows the local structure surrounding Co dimer. The gray plane is formed by three oxygen atoms (Red balls) O(1), O(2) and O(3). The center of the gray plane denoted as ``X'' has no inversion symmetry, therefore, DM interaction allows between Co(1) and Co(2). There are three  superexchange mediated DM interactions formed by Co(1) - O(1) - Co(2), Co(1) - O(2) - Co(2), and Co(1) - O(3) - Co(2). The direction of the DM interaction of each bond is denoted as yellow arrow at oxygen atoms. These bonds are connected by C$_{3}$ rotational symmetry along {\it c}-axis, which makes them cancel each other to result in net zero DM interaction. 

\bibliography{main}

\end{document}